\documentclass[twocolumn,trackchanges]{aastex631}
\usepackage[utf8]{inputenc}
\usepackage{graphicx}
\usepackage{longtable}
\usepackage{verbatim}
\usepackage{array,multirow,makecell}
\usepackage{amsmath}
\usepackage{siunitx}
\DeclareSIUnit\au{au}
\DeclareSIUnit\AccrRate{\ensuremath{M_{\astrosun}\text{yr}^{-1}}}
\usepackage{wasysym}
\usepackage{booktabs}
\usepackage{svg}
\usepackage[para, online]{threeparttable}
\usepackage[version=4]{mhchem}

\submitjournal{ApJL}

\shorttitle{Water in the Proto Solar Nebula}
\shortauthors{Boitard-Crepeau et al.}

\begin{document}

%\title{Origin of Water on Earth: a Diffuse Snowline Model from Computational Chemistry}
%%%CC: I think that this title is not appealing, not informative (who cares abuout computational chemistry, this is vague to understand what the article is about) and obscure, because nobody knows what a "diffuse snowline" is... so I try the following different titles:
%\title{Is a delivery from material at $\geq$ 1 au necessary to explain the terrestrial water origin?}
% or
%\title{Was water on Earth acquired locally in the young Solar System?}
% or
\title{Was Earth's water acquired locally during the earliest phases of the Solar System formation?}

%%

%\correspondingauthor{August Muench}
\email{lise.boitard-crepeau@univ-grenoble-alpes.fr, cecilia.ceccarelli@univ-grenoble-alpes.fr}

\author[0009-0009-3735-0518]{Lise Boitard-Crépeau}
\affiliation{Univ. Grenoble Alpes, CNRS, IPAG, 38000 Grenoble, France\\}

\author[0000-0001-9664-6292]{Cecilia Ceccarelli}
\affiliation{Univ. Grenoble Alpes, CNRS, IPAG, 38000 Grenoble, France\\}

\author[0000-0002-6532-5602]{Pierre Beck}
\affiliation{Univ. Grenoble Alpes, CNRS, IPAG, 38000 Grenoble, France\\}

\author[0000-0003-4230-6748]{Lionel Vacher}
\affiliation{Univ. Grenoble Alpes, CNRS, IPAG, 38000 Grenoble, France\\}

\author[0000-0001-8886-9832]{Piero Ugliengo}
\affiliation{Dipartimento di Chimica and Nanostructured Interfaces and Surfaces (NIS) Centre, Universit\`{a} degli Studi di Torino, via P. Giuria 7, 10125, Torino, Italy\\}

\begin{abstract}
The origin of the terrestrial water remains debated, as standard Solar System formation models suggest that Earth formed from dry grains, inside the snowline of the Proto-Solar Nebula (PSN).  
Here, we revisit this issue through the lens of computational chemistry. 
While the classically used snowline relies on a single condensation temperature, recent work in quantum chemistry shows that the binding energy of water on icy grains has a gaussian distribution, which implies a gradual sublimation of water rather than a sharp transition. 
We use the computed distribution of binding energies to estimate the radial distribution of adsorbed ice on the dust grains across the PSN protoplanetary disk. 
Our model reproduces the full range of estimated water abundances on Earth and matches the hydration trends observed in chondrite groups at their predicted formation distances.
Thus, we suggest that a significant fraction of Earth’s water may have been acquired locally at early stages of the Solar System formation, without requiring delivery from beyond the classical snowline.
\end{abstract}


%%%%%%%%%%%%%%%%%%%%%%%%%%%%%%%%%%%%%%%%%%%%%%%%%%%%%%%%%%%%%%%%%%%%%%
%%%%%%%%%%%%%%%%%%%%%%%%%%%%%%%%%%%%%%%%%%%%%%%%%%%%%%%%%%%%%%%%%%%%%%
%%%%%%%%%%%%%%%%%%%%%%%%%%%%%%%%%%%%%%%%%%%%%%%%%%%%%%%%%%%%%%%%%%%%%%
\section{Introduction} \label{sec:introduction}

Earth's formation started 4.6 billion years ago, in the protoplanetary disk that gave birth to our Solar System, the so-called Proto-Solar Nebula (PSN). 
In the Solar System, water, or more generally "water-equivalent" for is H-bearing compounds (such as molecular water H$_2$O, hydroxyl OH, and molecular hydrogen H$_2$), is found throughout all kind of objects, %accounting for $\sim$ 50\% of the mass of all condensable species \citep{lodders2003}} \textit{water is the second most abundant molecule 
from the inner rocky planets and asteroids to the troposphere of the outer giant planets and their satellites, and in comets \citep[e.g.][]{Encrenaz2008-H2OSolarSystem, Peslier2022-Elements}. 
Yet its origin on Earth, whether acquired during its formation locally or elsewhere or even after Earth was mostly formed, is a hotly debated issue.

There is consensus on the fact that molecular water was mostly formed on micro-meter sized dust grains at the very beginning of the Solar System formation, in the molecular cloud and pre-stellar phase, where it remained frozen on the icy mantle enveloping the grains  \citep[e.g.][]{ceccarelli2022}. 
With time, these grains coagulated and grew, forming rocky planets, asteroids, comets and the other small bodies currently populating the Solar System. 
However, when the disk temperature increased with the formation of the Sun, water sublimated from the grain surfaces inwards the so-called snowline -the condensation front of water \citep[e.g.][]{hayashiPSN1981}. 
This caused the inner disk to get depleted of water that was not already trapped in larger bodies. 
Thus, standard Solar System formation models suggested that Earth accreted from the dry inner-disk material, implying that its water must have been delivered by sources outwards the Earth's orbit \citep[e.g.][]{morbidelli2000}. 
%CC: I removed the next sentence because is not important here and blur the message
%Other more recent models suggest that the snowline may have moved inwards the Earth's orbit during its formation \citep[e.g.][]{garaudLin2007, okaSL2011}, but this scenario would be hard to reconcile with the volatile‑poor nature of the inner Solar System.

The classical approach to define the snowline is based on a single condensation temperature of water, assumed to be 180 K \citep[e.g.,][]{lodders2003}. 
However, the sublimation of a frozen molecule critically depends on its adsorption strength on the grain surface, referred as the binding energy (BE), and this value is not unique as it depends on the different orientation of the molecules on different adsorbing sites of the surface. 
Relevant to the water snowline, recent theoretical quantum chemical calculations by \cite{tinacciBE2023} have shown that the water BE on an icy surface has a gaussian distribution, with a not negligible contribution of BE with values larger than that usually adopted in the PSN models. 
These authors also showed that the water sublimation on a generic protoplanetary disk occurs over an extended region, when considering their computed BE distribution.
In other words, the traditional concept of snowline, defined as an abrupt transition, is not appropriate and it would be more adequate to speak of a “water transition zone”.

In this work, we combine the \cite{tinacciBE2023} water BE distribution with the classical structure model of the PSN midplane to provide the theoretical radial distribution of the water ice, with respect to the dust mass, across the PSN disk midplane (Sec. \ref{sec:model}).
We then present a selection of a sample of chondrites, thought to be remnants of the early Solar System, and review their water content along with that of Earth (Sec. \ref{sec:Solar-System-data}) in order to compare the theoretical water abundance profile with estimates of the water contained in these objects (Sec. \ref{sec:results}).
We finally discuss the implications of this comparison (Sec. \ref{sec:discussion}) and conclude (Sec. \ref{sec:conclusions}).

%%%%%%%%%%%%%%%%%%%%%%%%%%%%%%%%%%%%%%%%%%%%%%%%%%%%%%%%%%%%%%%%%%%%%%
%%%%%%%%%%%%%%%%%%%%%%%%%%%%%%%%%%%%%%%%%%%%%%%%%%%%%%%%%%%%%%%%%%%%%%
%%%%%%%%%%%%%%%%%%%%%%%%%%%%%%%%%%%%%%%%%%%%%%%%%%%%%%%%%%%%%%%%%%%%%%

\section{Model description} \label{sec:model}

\subsection{PSN density and temperature profiles}

In this work, we adopt the classical Minimum Mass Solar Nebula (MMSN) recipe to describe the gas column density profile of the PSN midplane \citep[e.g.,][]{weidenschilling_1977}:
\begin{equation} \label{eq:PSNdens}
    \Sigma(r) = \Sigma_{1\text{au}} ~\left(\frac{r}{1 \ \text{au}}\right)^{-3/2} \ \si{\gram\per\centi\meter\squared},
\end{equation}
where $\Sigma_{1\text{au}}$ is the gas column density at 1 au distance, assumed to be \SI{1700} {\gram\per\centi\meter\squared} \citep{hayashiPSN1981}, and $r$ is the distance in au from the Sun.

Regarding the midplane temperature profile $T(r)$, we consider a passively heated disk, i.e. the solar radiation is the main heating source, and adopt the prescription developed by \cite{chiang_goldreich_1997}: 
\begin{equation} \label{eq:PSNtemp}
    T(r) = T_{1\text{au}} \left(\frac{r}{1 \ \text{au}}\right)^{-3/7} \ \text{K}.
\end{equation}
where $T_{1au}$ is the temperature at \SI{1}{\au}.
In the literature, $T_{1au}$ is usually assumed to be within 150--300 K, depending on the adopted Sun luminosity \citep{chiang_goldreich_1997, hayashiPSN1981, deschJupiter2018}. 
For example, \cite{kusaka1970} evaluated a temperature of 225 K at 1 au assuming a Sun luminosity of 10 L$_{\astrosun}$.
In our model, we treat $T_{1au}$ as a free parameter.

%%%%%%%%%%%%%%%%%%%%%%%%%%%%%%%%%%%%%%%%%%%%%%%
\subsection{Water chemistry and snowline}

We assume that water is mostly in the ice formed during the prestellar core phase \citep[e.g.][]{Ceccarelli2014-PP6}.
Then, at each radius $r$, its abundance x(H$_2$O) is given by the equilibrium of the adsorption of gaseous water onto the grain surfaces and thermal desorption of frozen water\footnote{As the midplane of the disk is optically thick to UV photons, water photodesorption and photodissociation can be neglected.}.

The adsorption rate of gaseous water onto the dust grains $k_{ads}$ is described by the usual equation:
\begin{align}
    k_{ads} & = S_{gr} ~\sigma_{gr} ~n_{gr} ~v_{th},
\end{align}
where $S_{gr}$ is the sticking coefficient, here taken equal to 1, $\sigma_{gr}$ is the geometrical cross section of a 0.1 $\mu$m grain\footnote{We assume the initial dust grain radius equal to the average one in the Inter-Stellar Medium.}, $n_{gr}$ is the grain number density, computed using the standard mass gas to dust ratio of 0.01, and $v_{th}$ is the water thermal velocity equal to $\sqrt{2~k_B ~T ~/~ m_{H_2O}}$. 

The thermal desorption rate of frozen water $k_{des}$ is given by the Polanyi-Wigner equation, first order \citep{Minissale2022, Ceccarelli2023-PP7}:
\begin{equation}
    k_{des} ~=~ \nu ~\exp\left(-\frac{BE}{T}\right)
\end{equation}
where $\nu$ is the pre-exponential factor, $BE$ is the water binding energy (in K), and $T$ the temperature of the grain surface, assumed to be equal to that of the gas\footnote{Gas and dust are thermally coupled at the densities of the PSN disk midplane, as the coupling is efficient at densities larger than about $10^5$ cm$^{-3}$ \citep[e.g.,][]{Goldsmith2001}.}. 
Here, we use the water BE computed by \cite{tinacciBE2023}, which showed that it has a Gaussian-like distribution. 
Specifically, they identified 10 different BE bins, with BE ranging from 14.2 to 61.6 kJ mol$^{-1}$, here shown in Fig. \ref{fig:BEsnowlines}, upper right panel.  

When all BE sites of the adsorbate surface are occupied by the adsorbed species, as it is the case for water which is the adsorbate surface and the desorbing species at the same time, the desorption-adsorption equilibrium can be computed by considering it for each separate BE bin and it holds:
\begin{gather}
    k_{ads} ~n_{H_2O}^i ~=~ k_{des}^i ~n_{ice}^i \label{eq:chem1}\\
    n_{H_2O}^i ~+~ n_{ice}^i ~=~ f_{ice}(BE_i) ~A_{ox} ~n_{H} \label{eq:chem2}\\
    \sum_i n_{ice}^i ~+~ \sum_i n_{H_2O}^i ~=~  A_{ox} ~n_{H} \label{eq:chem3}
\end{gather}
where $n_{H_2O^i}$, $n_{ice}^i$ and $f_{ice}(BE_i)$ are respectively the number densities of gaseous water, the frozen water (ice) on the grains and the fraction of occupied adsorption sites at each BE bin $i$ (i.e., the "fraction of the ice" in the upper right panel of Fig. \ref{fig:BEsnowlines}); $A_{ox}$ is the total water abundance and $n_{H}$ is the total H-nuclei density respectively.

In our model $A_{ox}$ is a parameter.
Assuming that all the oxygen not trapped in the refractory component of the dust grains, nor in CO, is locked into water, we adopt $A_{ox} ~=~ 2\times 10^{-4}$ (with respect to H-nuclei). 
This value is derived from the amount of water, with respect to the refractory component, measured in the comet 67P/Churyumov-Gerasimenko (hereafter referred as 67P) (see Sec. \ref{sec:Solar-System-data}), considered to be a pristine remnant of the early Solar System \citep[e.g.,][]{choukrounChury2020}. 

Finally, we solve Eqs. \ref{eq:chem1} to \ref{eq:chem3} at each radius $r$ of the PSN, using the density and temperature profiles of Eqs. \ref{eq:PSNdens} and \ref{eq:PSNtemp}.

%%%%%%%%%%%%%%%%%%%%%%%%%%%%%%%%%%%%%%%%%%%%%%%%%%%%%%%%%%%%%%%%%%%%%%
%%%%%%%%%%%%%%%%%%%%%%%%%%%%%%%%%%%%%%%%%%%%%%%%%%%%%%%%%%%%%%%%%%%%%
%%%%%%%%%%%%%%%%%%%%%%%%%%%%%%%%%%%%%%%%%%%%%%%%%%%%%%%%%%%%%%%%%%%%%%

\section{Solar System data} \label{sec:Solar-System-data}

Our simple model pictures a protoplanetary disk made of gas and dust grains covered by icy mantles that progressively gets thinner when getting closer to the Sun. 
As the disk evolves, grains coagulate to form planetesimals, which will then accrete more materials to form the planets. 
During this process, water is mixed with the dust and trapped in the solid bodies. 
Asteroids and comets are the remnants of this initial population of planetesimals and are, thus, the most pristine record we have of the early composition of the Solar System. 

The chemical and isotopic analysis of meteorites, fragments of asteroids recovered from their fall on Earth, provide strong constrains on the formation history of the different planetesimals/planets. 
On the other hand, meteorites can suffer from terrestrial weathering on a very short timescale and samples returned through space mission, like Hayabusa2 \citep{yadaHayabusa22022} and OSIRIS-REx \citep{laurettaOSIRISRExSampleReturn2017} sent to Ryugu and Bennu respectively, are particularly pristine and valuable. 
Rosetta's flyby and touchdown at the comet 67P also provided prime data on the composition of the outer Solar System \citep{glassmeierRosetta2007}.

In the following, we briefly review the water-equivalent content of chondrites through their measured amount of hydrogen (Sec. \ref{subsec:chondrite-sample}) and their position at their formation.

\begin{table*}
\centering
\begin{threeparttable}
\caption{Water-equivalent content by weight of different chondrites and their estimated formation position.
First two columns identify the object; 
third column quotes the water in percentage contained in the mineral matrix; columns 4--7 list the median, minimum and maximum water content in mass, and the references (see Sec. \ref{subsec:chondrite-sample}); 
last three columns provide the radius at which each object formed with the possible minimum and maxim values (see Sec. \ref{subsec:chondrite-position}).}
\label{tab:chondrites}
\begin{tabular}{llcrrrlccc}
\toprule
                            &                                              &     Matrix                   & \multicolumn{4}{c}{H$_2$O eq. \tnote{$\dagger$} (weight \%) }                           & \multicolumn{3}{c}{Position (au)}   \\ 
    &type                   & \shortstack{(weight \%)}& Median                & Min       & Max  & Ref.                   & $r$ & $r_{min}$   &$r_{max}$ \\
\hline
Comet 67P  &               & 100      & 18. & 12. & 25. & 1           & 25 & 20 & 50\\ 
\hline
CC & CI \tnote{$\star$}    & 100      & 7.6 & 5.1 & 11. & 2,3,4,5,6,7 & 15 & 10 & 20\\
   & CR                    &  30      & 14. & 4.3 & 20. & 4,6,8       & 10 &  3 & 10\\ 
   & CM                    &  60      & 11. & 6.1 & 14. & 3,4,6       &  7 &  3 & 10\\ 
   & CO                    &  30      & 11. & 9.0 & 14. & 9,10,11     &  5 &  3 & 10\\ 
\hline
NC & Ordinary              & 12       & 5.7 & 2.6 & 12. & 10,11       &  2 & 1.5& 3\\ 
   & Enstatites            &          & 0.34& 0.22& 0.43& 12          & 1.25&1.0& 1.5\\ 
\hline
 Earth  &                  &          & 0.39& 0.04& 3.6 & 13          & 1   &   & \\ 
\bottomrule
\end{tabular}
$\dagger$ water-equivalent content in weight \% normalised to matrix proportion $^*$ including Ryugu and Bennu.
\textit{References:} 
    1- \cite{choukrounChury2020}; 2- \cite{yokoyama2023a}; 3- \cite{vacher2020}; 4- \cite{kerridge1985}; 5- \cite{robertEpstein1982}; 6- \cite{kolodny1980}; 7- \cite{laurettaBennu2024}; 8- \cite{alexanderCMCR2013}; 9- \cite{alexanderCO2018}; 10- \cite{grantBulkMineralogyWater2023}; 11- \cite{vacherOC2024}; 12- \cite{pianiEnstatite2020}; 13- \cite{peslierWaterEarth2017}.
\end{threeparttable}
\end{table*}

%%%%%%%%%%%%%%%
\subsection{Chondrite samples selection and water content} \label{subsec:chondrite-sample}

Meteorites are generally classified in two major groups based mainly on their bulk composition and textures \citep{krot2014}: achondrites (differentiated meteorites) and chondrites (undifferentiated). 
%Since achondrites are igneous stony and iron meteorites, resulting from internal melting of their parent bodies, they have lost their volatiles and are thus not of interest here. 
Chondrites have the potential to preserve their primitive constituents, and are of prime interest for the study of the history of the Solar System. 

An isotopic dichotomy has been recorded within the chondrites \citep{trinquier2007, warrenDichotomie2011}, separating the Carbonaceous Chondrites (CC) from the non-carbonaceous chondrites (NC). 
The CC are also enriched in matrix and volatiles, and are subdivided in several groups according to their composition (like CI, CM, CO, CR...). 
The NC are parted in two main classes: the ordinary chondrites (OC) and the enstatites chondrites (EC). 
After accretion, chondrite parent bodies underwent geological processes, in particular thermal metamorphism (that dehydrates the minerals), as well as aqueous alteration. 
During aqueous alteration, water ice accreted by the parent bodies melted because of the radioactive decay of $^{26}$Al heat \citep{leeAqueousAlteration2025}. 
The resulting aqueous fluid chemically interacted with the anhydrous silicates, leading to the incorporation of molecular water into silicate minerals, mostly in the form of hydroxyl group -OH in the structure of hydrous minerals, or as hydrogen bonded to oxygen atoms at lattice defects in anhydrous minerals \citep{peslierWaterEarth2017}. 
The bulk water-equivalent content measured in aqueously altered chondrites is, thus, related to the amount of water-ice initially accreted, and can provide constraints on the source of water on Earth.
% Since thermal metamorphism dehydrates chondrites, only low thermally metamorphosed (petrologic type PT 3) chondrites and aqueously altered chondrites (PT 1 and 2) are the most relevant meteorites to study the history of water in the Solar System. }

\cite{vacher2020} demonstrated the importance of pre-degassing the chondrite samples to analyse their hydrogen content, as this allows the desorption of weakly bonded \ce{H2O} sensitive to terrestrial contamination \citep{garenneTGA2014}.  
In order to build up a sample of chondrites to compare with the Sec. \ref{sec:model} model predictions, we selected studies that mentioned a pre-degassing step, and limited ourselves to chondrites that experienced aqueous alteration (petrologic type PT 1 and 2) and little thermal metamorphism ( PT $\leq$ 3.2), with a limited weathering grade ($\leq$ W2 or grade B). 
%Hence, we report here the measured amount of water in EC \citep{pianiEnstatite2020}, OC (with PT$\leq$3.2) \citep{grantBulkMineralogyWater2023, vacherOC2024}, CO \citep{alexanderCO2018}, CM \citep{kolodny1980, kerridge1985, eilerkitchen2004, vacher2020}, CR \citep{kolodny1980,kerridge1985, alexanderCMCR2013} and CI \citep{kolodny1980, robertEpstein1982, kerridge1985, eilerkitchen2004, kingCI2015, vacher2020}.
Hence, we report here the measured amount of hydrogen in EC, OC (both with PT$\leq$3.2), CO, CM, CR and CI, expressed as water-equivalent content by weight (see Tab. \ref{tab:chondrites} and references there).
Sample returns from the asteroids Ryugu \citep{yokoyama2023a} and Bennu \citep{laurettaBennu2024}, and from the comet 67P \citep{choukrounChury2020} were also considered.

Since hydrogen (contained in water equivalent compounds) is mainly accreted in the matrix of the chondrites \citep{garenneTGA2014}, all reported data have been systematically recalculated using the amount of matrix (weight \%)   measured by \cite{alexanderCC2019} for the CC and \cite{grossmanOC2005} for the OC (Tab. \ref{tab:chondrites}).
Here we assume that the bulk hydrogen in chondrites mainly comes from hydrous minerals, which is a reasonable hypothesis for CC regarding the low amount of hydrogen stored in organic matter compared to hydrous minerals \citep{alexanderScience2012}. For example, \cite{alexanderOrganicMatter2017} reported an elemental composition of the organic matter in chondrite as \ce{C100H_{70-80}O_{15-20}N_{3-4}S_{1-4}}.
With a maximum carbon content of 4 wt\% \citep{alexanderScience2012} in CC, the amount of hydrogen stored in the organic matter would be less than 0.3 wt\% for CC, corresponding to 0.9 wt\% of water-equivalent, which fall in the range of uncertainties of the measured bulk content.
In EC and OC however, not all the hydrogen are structurally bounded in silicate minerals. For example, $\sim$20 to 65\% of the total hydrogen measured in bulk EC or OC can be stored either in chondrules mesostasis, in hydrogen bonded to sulfur (H$_2$S), and/or in organic matter \citep{pianiEnstatite2020, barrettSourceHydrogenEarths2025, shimizuHighlyVolatileElement2021}. Since these reservoirs are not well quantified, we consider the bulk hydrogen concentration of EC and OC as upper limits on the amount of hydrogen derived from molecular water.  Table \ref{tab:chondrites} summarizes the adopted amount of water-equivalent content in each class of chondrites.

% Since water has mainly been accreted in the matrix of the chondrite \citep{garenneTGA2014}, all reported data have been systematically recalculated using the amount of matrix (weight \%)   measured by \cite{alexanderCC2019} for the CC and \cite{grossmanOC2005} for the OC, reported in Table \ref{tab:chondrites}. The amount of water is referred as \textit{water-equivalent content} by weight, calculated from the hydrogen content measured in bulk chondrites. We here assume that this hydrogen comes mainly from hydrous minerals in types 1 and 2 chondrites, which is a reasonable hypothesis regarding the low amount of organic matter in the chondrite.  \cite{alexanderOrganicMatter2017} reported an elemental composition of the organic matter in chondrite as \ce{C100H_{70-80}O_{15-20}N_{3-4}S_{1-4}}. With a maximum of carbon content of 4 wt\% \cite{alexanderScience2012} in CI for carbonaceous chondrites and less than 1 wt\% in  OC \cite{vacherOC2024}, the amount of hydrogen stored in the organic matter would be less than 0.3 wt\% for CC and less than 0.07 wt\% for OC. These values fall in the range of uncertainties for the measured hydrogen bulk contents.  In EC however, hydrogen from the chondrules and the organic matter can account for 20 \% of the total H measured \citep{pianiEnstatite2020}, the data set for EC when so corrected accordingly.

%%%%%%%%%%%%%%%%%%%%%%%%%%%%%%%%%%%%%%%%%%%%%%%
\subsection{Chondrites formation position} \label{subsec:chondrite-position}

%%% CC: I GREATLY REDUCED THE INITIAL TEXT, MOVED AT THE END OF THE DOCUMENT IN CASE WE WANT TO REINSERT SOME PARTS, BECAUS EHERE WE ONLY NEED TO SAY WHERE THE CC OF OUR SAMPLE ARE LIKELY FORMED -ALL THE HISTORY ON JUPITER ETC SEEMS TO ME NOT RELEVANT IN THIS CONTEXT
In order to compare the water-equivalent content of chondrites with our model, we need an estimate of where their parent bodies formed.  
While this remains debated, particularly for CC, the isotopic analysis can help constrain their possible origins.
For example, \cite{trinquier2007} established an isotopic dichotomy between the CC and NC, indicating the formation of their parent bodies in distinct reservoirs. 
Furthermore, various models suggest the mixing of the reservoirs after the formation of Jupiter, leading to the current structure of the asteroid belt.

In the following, we consider the study by \cite{demeoNature2014}, who presented a taxonomic distribution of the asteroids, where parent bodies of the OC populate the inner main-belt. CO are distributed towards the outer main belt, as well as asteroids analogues to CM chondrites. 
Objects that are spectroscopically similar to Tagish-Lake meteorites have been proposed to be present among Jupiter-trojans. 
The locations of CI-chondrites parent bodies remain debated. While the current location of CC-like asteroids seems to be mostly in the asteroid main-belt, their original formation location may have been more distant from the Sun \citep[see][and references therein]{demeoNature2014}, explaining the range of heliocentric distance quoted in Table \ref{tab:chondrites}.
Using these constrains, we assign an approximate birthplaces for each class of chondrites: 
ECs  between the Earth and Mars; OC and CC distributed from the Asteroid Belt to the outer Solar System in the following order: CO, CM, CR and CI. 
Table \ref{tab:chondrites} summarizes the adopted position for each class of chondrites.

In addition, we assumed that the comet 67P was formed in the outer Solar System, at \SI{25}{\au}, following the arguments by \cite{marschall2025}. 

Please note that the adopted formation regions provide a useful framework for the present analysis; we do not, however, claim to provide a precise reconstruction of the parent‑body formation sites here.

\begin{figure*}
    \centering
    \includegraphics[width=\textwidth]{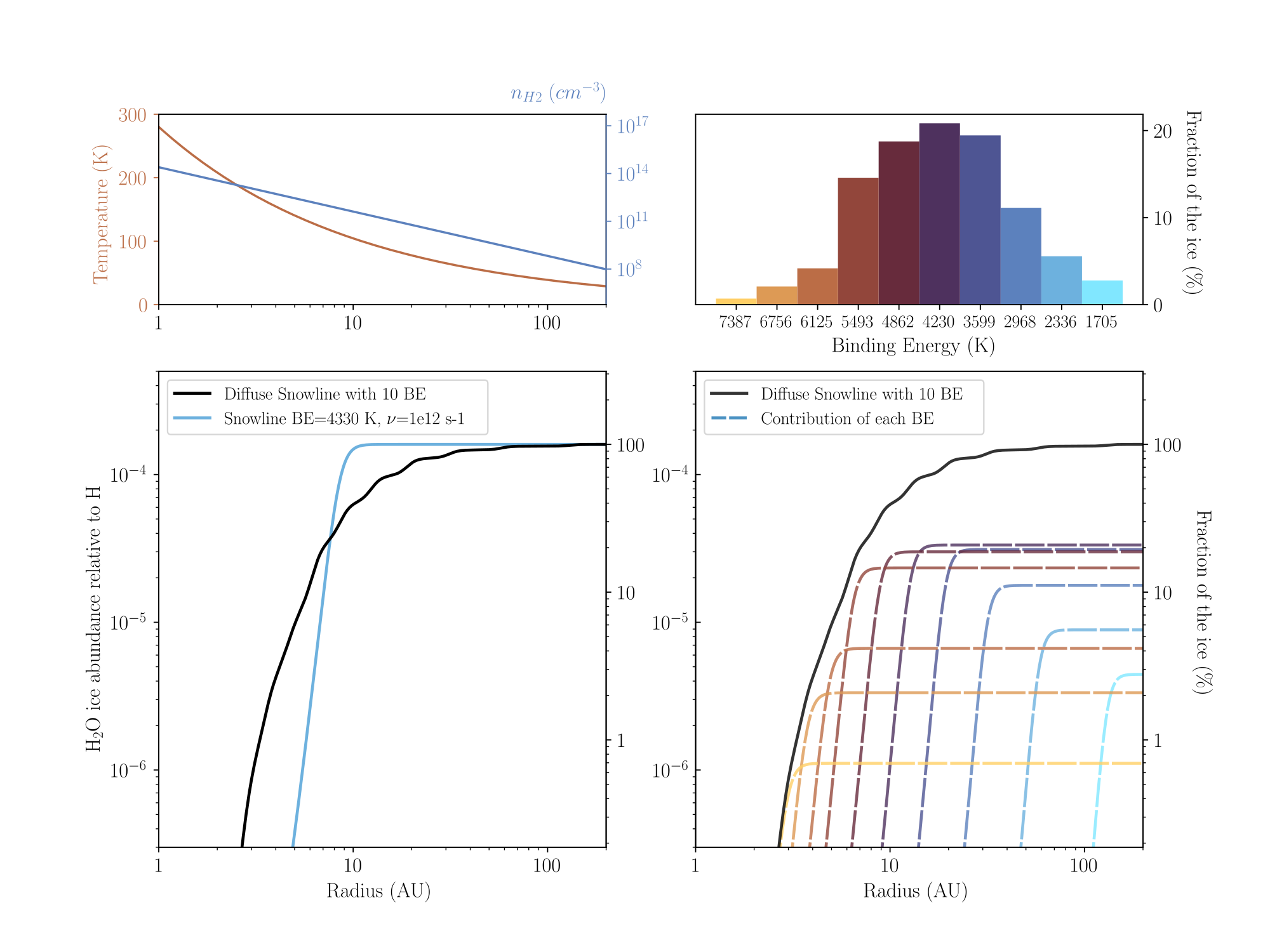}
    % \includesvg[width=\textwidth]{BEsnowlines}
    \vspace{-1 cm}
    \caption{Temperature and density profiles, snowlines and contribution of each BE.\\
    \textit{Upper left:} Temperature (brown) and density (blue) profile of the midplane with $T_{1au}$ = 280 K in Eq. \ref{eq:PSNtemp}.\\
    \textit{Upper right:} Distribution of the water BEs distribution computed by \cite{tinacciBE2023}.\\
    \textit{Lower left:} Comparison between a snowline computed with a single BE \citep[BE = 4330 K and $\nu = \SI{1e12} {\per\second}$;][]{collingsBE2015}) (cyan) and using the \citep{tinacciBE2023} water BE distribution (black).\\ 
    \textit{Lower right:} The dashed curves show the contribution of each BE bin with the matching colour from the \cite{tinacciBE2023} distribution above. The black curve is the same as the one on the left panel and uses the whole BE distribution. \\
    The left and right lower panels share the same vertical axis: water ice abundance relative to H-nuclei on the left axis and the amount of ice relative the maximum on the right axis ($x_{ice} / x_{ice, max}$ in \% ). }
    \label{fig:BEsnowlines}
\end{figure*}

%%%%%%%%%%%%%%%%%%%%%%%%%%%%%%%%%%%%%%%%%%%%%%%
\newpage
\subsection{Water on Earth} \label{subsec:earth-water}

Atmospheric and oceans water only represent \SI{236} {ppm} of the Earth's total mass, according to the estimates obtained by \cite{peslierWaterEarth2017}. 
Actually, Earth's water is mainly present as hydrogen incorporated into the minerals of its crust, mantle and core. 
The total amount of hydrogen is, however, not well constrained, especially in the core, leading to large uncertainties, up to a factor 10.
The most recent estimates of the total terrestrial water-equivalent content is situated at $3900^{+32700}_{-3300}$ ppm \citep{peslierWaterEarth2017}.

\begin{figure*}
    \centering
    \includegraphics[width=\linewidth]{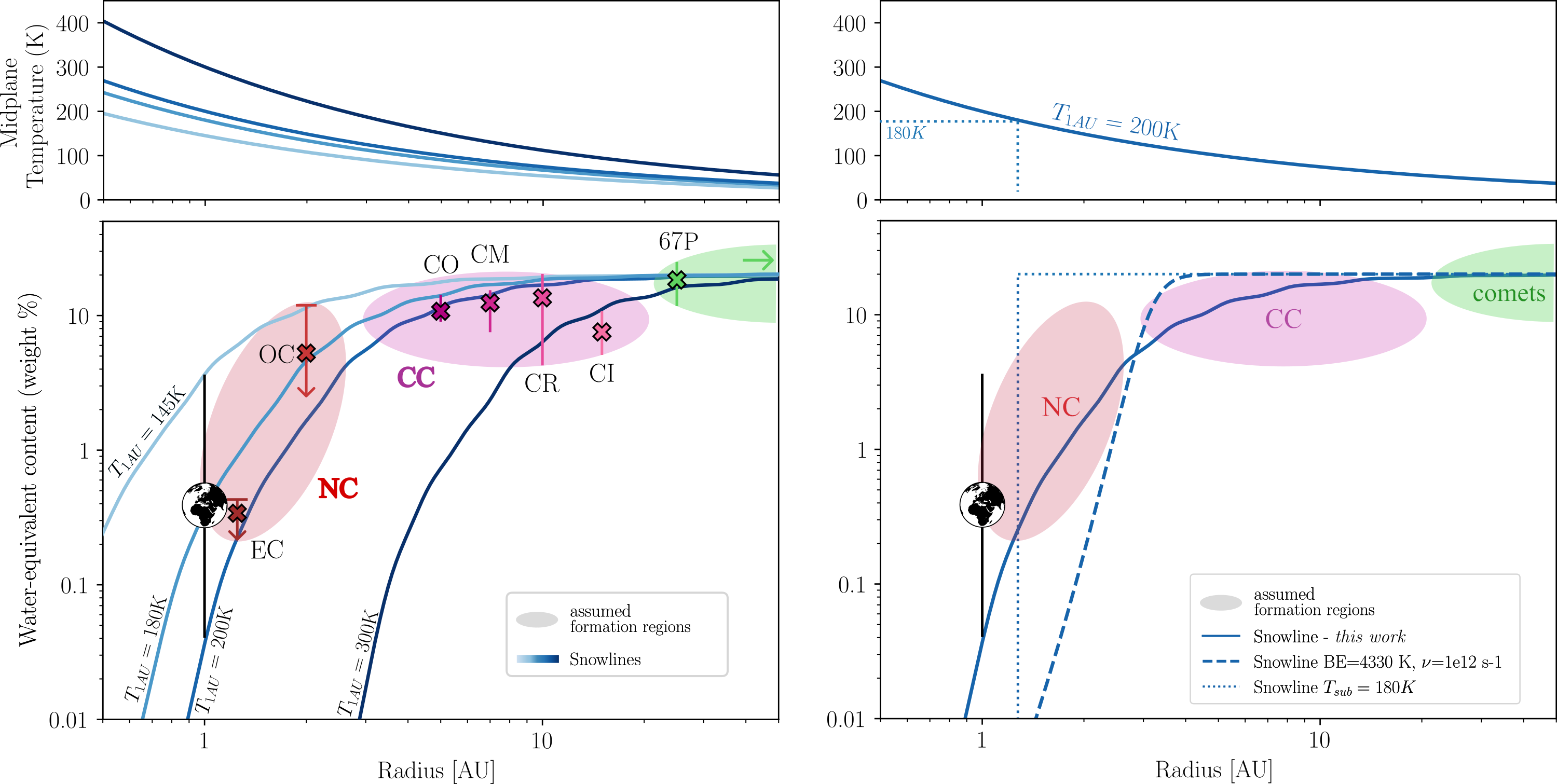}
    % \includesvg[width=\linewidth]{Snowlines_Meteorites}
    \caption{Water-equivalent content by weight in the Solar System and model predictions.\\ 
    \textit{Left upper panel:} Midplane temperature of the disk using $T_{1au}$ = 145, 180, 200 and 300 K (from light blue to dark blue respectively) in Eq. \ref{eq:PSNtemp}.\\
    \textit{Left lower panel:} The solid curves show the snowlines calculated using the \cite{tinacciBE2023} BE distribution for the different temperature profiles (colours code as above).
    The data points show the median values of the water-equivalent content by weight  for different objects of the Solar System (Sec. \ref{sec:Solar-System-data}), where vertical bars represent the minimum and maximum range.
    Please note that the values reported in Tab. \ref{tab:chondrites} are considered upper limits (see text).
    The uncertainty of the formation region of the various objects are shown as ellipses, distinguishing NC, CC, and comet populations. \\
    \textit{Right:} Computed snowlines using $T_{1au}$ = 200 K in Eq. \ref{eq:PSNtemp} and different models (see text): BE distribution (solid line), single BE (dashed) and condensation at 180 K (dotted). \\
    The right and left panels share the same vertical axis.}
    
    \label{fig:SnowLinesMeterorites}
\end{figure*}

%%%%%%%%%%%%%%%%%%%%%%%%%%%%%%%%%%%%%%%%%%%%%%%%%%%%%%%%%%%%%%%%%%%%%%
%%%%%%%%%%%%%%%%%%%%%%%%%%%%%%%%%%%%%%%%%%%%%%%%%%%%%%%%%%%%%%%%%%%%%%
%%%%%%%%%%%%%%%%%%%%%%%%%%%%%%%%%%%%%%%%%%%%%%%%%%%%%%%%%%%%%%%%%%%%%%
\section{Results} \label{sec:results} \label{sec:results}

We first discuss the predicted ice abundance as a function of the radius $r$ assuming $T_{1au} = \SI{280} {\kelvin}$ in Eq. \ref{eq:PSNtemp}, following \cite{hayashiPSN1981}, and show the impact of introducing a BE distribution of values rather than a single one.
We then vary $T_{1au}$ in order to find the ones that best reproduce the measured water-equivalent content in the chondrite sample, the comet 67P and Earth.

\paragraph{$T_{1au}$ = 280 K}
The upper panels of Fig. \ref{fig:BEsnowlines} show the temperature and density profiles of the PSN disk midplane and the BE distribution computed by \cite{tinacciBE2023}.
With this structure, we first compute the snowline obtained using a single value, 
that experimentally measured by \cite{collingsBE2015}: 
%%% CEC: but we compare orgages with apples here! our BE are on amorphous water ice, I just removed, as, afetr all, it is not important
%on amorphous silicates
\SI{4330} {\kelvin} and $\nu$ = \SI{1e12} {\per\second}.
The lower left panel of Fig. \ref{fig:BEsnowlines} shows that, in this case, the transition ice-vapour is very sharp: water ice starts to desorb around 10 au and at $\leq$4 au only $\leq$0.2\% remains frozen. 

On the contrary, using the \citep{tinacciBE2023} BE distribution spreads the water ice sublimation front over a much wider region.
The ice desorption starts already around 150 au but it is much smoother: for example, at 4 au about 6\% of water remains frozen, namely more than a factor 30 larger than when using the \cite{collingsBE2015} BE.
The lower right panel of Fig. \ref{fig:BEsnowlines} show the individual contribution of each BE bin, highlighting the progressive sublimation of the ice. 
The least strongly bonded water molecules (2.7\% of the ice) start to desorb around 150 au, but the 0.07\% most strongly bonded ones only desorb at  2 au. 

\paragraph{$T_{1au}$ variation}
We ran the model varying $T_{1au}$ from 145 to 300 K, to explore the range of $T_{1au}$ evoked in the literature.
The left panels of Fig. \ref{fig:SnowLinesMeterorites} show the temperature profiles (upper panel) and the associated snowlines (lower panel).
In the same figure, we also reported the water-equivalent content by weight  of each type of chondrite at their estimated formation position, along with the amount of water measured in the comet 67P and on Earth. 
The comparison between the theoretical ice distribution and the water content of the Solar System objects shows that a structure with a temperature profile having between 145 and 200 K at 1 au would reproduce, within the error bars, all the measured data.

To show the dramatic difference when considering a BE distribution rather than a single value, the right panels of Fig. \ref{fig:SnowLinesMeterorites} show the snowlines obtained using $T_{1au}$ = 200 K in the two cases.
Using a single BE value fails to reproduce the Earth's water by several orders of magnitude and just barely reproduce that of a few NC chondrites.
Finally, the same figure shows also the condensation snowline (see Introduction) of 180 K, similarly unable to reproduce the Earth's water and NC chondrites.

%%%%%%%%%%%%%%%%%%%%%%%%%%%%%%%%%%%%%%%%%%%%%%%%%%%%%%%%%%%%%%%%%%%%%%
%%%%%%%%%%%%%%%%%%%%%%%%%%%%%%%%%%%%%%%%%%%%%%%%%%%%%%%%%%%%%%%%%%%%%%
%%%%%%%%%%%%%%%%%%%%%%%%%%%%%%%%%%%%%%%%%%%%%%%%%%%%%%%%%%%%%%%%%%%%%%
\newpage
\section{Discussion} \label{sec:discussion} \label{sec:discussion}

Our modelling using a distribution of the molecular water BEs rather than a single value shows that enough water can be retained by the dust grains that eventually formed the Earth, if they mostly coagulated when their temperature was around 145--200 K at 1 au (Fig. \ref{fig:SnowLinesMeterorites}).
The temperature 145 and 200 K are necessary if the Earth's water is the upper and lower limits of the \cite{peslierWaterEarth2017} estimate, respectively. 
In other words, the adoption of a water BE distribution does not generate a single snowline, inside which water ices are fully desorbed and remain "completely dry", as often assumed, but rather a water transition zone extending several au.
Models simulating the evolution of the PSN, for example, often consider that the snowline lies at about 5 au following this criterium  \citep{Morbidelli2022-PSNmodel, morbidelli2025}.
However, while the bulk of ice is indeed desorbed at larger distances than 1 au, a small fraction remains attached to the grain surfaces inwards (Fig. \ref{fig:BEsnowlines}).
This small fraction, between $\sim$ 0.04 and 2.5 weight \%, can fully account for the Earth's water content, depending on the \cite{peslierWaterEarth2017} estimate.
In turn, terrestrial water could be mostly inherited from the dust grains that were in the Earth's orbit, with no necessity of migration of outer ones. 

Our model also successfully reproduces the observed water-equivalent content trend across chondrite groups.
While their parent bodies accreted at different times\footnote{This is demonstrated by radiometric analysis of CAIs (Calcium-Aluminium-rich Inclusions), chondrules, and secondary minerals, \citep[e.g.][]{sugiuraAccretion2014b,spitzer2020}.}, their measured water-equivalent contents likely represents lower limits of primitive water originally incorporated into silicate dust grains, the building blocks of chondrite matrices. 
%%% CC: I'M NOT SURE I UNDERSTAND, I REMOVE FOR THE MOMENT
%This suggests that even with temperature as high as \SI{200} {\kelvin}, early grain growth, for example in Class 0 protostars, could have trapped sufficient water to be later processed during chondrite formation.

There are still a few questions that need to be answered to make our claim of local inheritance robust: 
(1) Is a PSN temperature of 145--200 K at 1 au reasonable and, if yes, when in its evolution?\\
(2) Is the hypothesis of a major dust grains coagulation up to those temperatures reasonable?\\
(3) Is the hypothesis of a fully local inheritance compatible with the measured isotopic water ratio HDO/H$_2$O?\\
In the following, we will provide arguments to answer these questions, based on the observations of solar-like planetary systems currently forming in our Galaxy.
Following the mounting evidence that dust grain coagulation and possibly planetesimal formation starts in the Class 0/I protostellar phases \citep[e.g.,][]{Miotello2014, Han2023-GrainGrowth, HuEarlyGrainGrowth2025}, when the disk where planets form is still embedded in its placental envelope, we will focus on those two phases.

\paragraph{(1)} \textit{Is a PSN temperature of 145--200 K at 1 au reasonable and, if yes, when in its evolution?}\\
There are not direct observations of dust temperatures at $\sim 1$ au scales, because at a typical distance of 100--200 pc it would need a IR telescope with a spatial resolution of $\sim$0.005--0.01 arcsec.
To our best knowledge, \cite{Frediani2025-irasa4Temp} reported the measurement of the gas temperature at small scales in a Class 0 protostar using the ALMA interferometer.
These authors measured a temperature of 120--400 K between 19 and 43 au with the temperature dependence on the radius of $\sim 0.8\pm0.2$ power law (their Fig. 5).
The measurements were obtained towards the Class 0 protostar NGC1333 IRAS4A, which has a luminosity of $\sim 15 \pm 3$ L$_\odot$ when its distance of 293 pc is taken into account \citep{Zucker2018}.
Scaling the measured temperature at $22\pm3$ au of $250^{+160}_{-70}$ K to a distance of 1 au and a luminosity of 0.5 L$_\odot$ would give a temperature lower limit of 170 K, within the values derived by our analysis above (Sec. \ref{sec:results}).
In practice, temperatures of 145--200 K at 1 au would be possible when the young Sun was a Class 0 protostar with only a luminosity of 0.5 L$_\odot$.

\paragraph{(2)}\textit{Is the hypothesis of a major dust grains coagulation up to those temperatures reasonable?}
Again, measurements of the grain sizes during the early evolution of solar-like planetary systems are very scarce and indirectly derived by modelling of the dust emission.
At present, the system HH212-mm is the best studied. 
Recent work by \cite{HuEarlyGrainGrowth2025} corroborates the findings of \cite{leeHH2122021}, revealing early grain growth by more than three orders of magnitude in this Class 0 protostellar disk. 
This suggests that water could become trapped inside growing dust grains during the early stages of disk evolution. 
Subsequent heating events may sublimate water at the grain surfaces, while the iced trapped inside would remain protected.

\paragraph{(3)} \textit{Is the hypothesis of a fully local inheritance compatible with the measured isotopic water ratio HDO/H$_2$O?}
The Earth [HDO]/[H$_2$O] ratio, $3.1\times 10^{-4}$  \citep[the VSMOW value;][]{lecuyerHydrogenIsotopeComposition1998}, equivalent to an elemental D/H ratio of $1.56\times 10^{-4}$.
However, this value may not represent the original (primordial) D/H of the terrestrial water bulk, because of various processes that cycles water between the surface and the Earth's interior \citep{Hallis2017-DinSolarSystem}.
Nonetheless, measurements of the D/H ratio in the Earth's deep mantle provide an about 20\% lower value \citep{Hallis2015S-DinDeepMantle, Hallis2017-DinSolarSystem}.
Likewise, the distribution of D/H in CCs is peaked around the VSMOW value, with a shoulder extending up to a factor $\sim1.5$ larger values \citep[e.g., Fig. 9 of][and reference there]{Ceccarelli2014-PP6}.
Although the D/H ratios are not exactly the same in chondrites and the Earth's water, they are significantly (around a factor 10) enriched in deuterium compared to the PSN,  which has an elemental D/H ratio roughly 10 times lower \citep[$2.1\times 10^{-5}$][]{Geiss1998-PSNdeuterium}.
It has long been known that large molecular deuteration ratios are a hallmark of cold formation conditions \citep[e.g.,][]{Ceccarelli2014-PP6, Nomura2023-PP7}.
Observations in Class 0/I protostars provide measurements of the [HDO]/[H$_2$O] abundance ratio of 2--20$\times 10^{-4}$ \citep{Taquet2013-obs, Persson2014-HDO, Jensen2019-HDO, Andreu2023}, which is a D enrichment up to one order of magnitude larger than that observed on Earth and in both CCs and NCs \citep[e.g.][]{pianiizidoroOriginWaterTerrestrial2022}.

We emphasise that the 2--20$\times 10^{-4}$ values are larger than those measured in the Earth, chondrites and comets, so it is worth to understand whether they are compatible with the latter.
We notice, however, that they refer to the totality of the water ices sublimated in the hot corinos, the regions where the dust temperature is larger than the sublimation temperature of the ice bulk.
Likewise, the recent detections of frozen HDO around young stars yield ratios of $\sim 10^{-3}$ \citep{Slavicinska2024-HDOsolid}. 
It is important to note here that both types of observations, by definition, provide the [HDO]/[H$_2$O] averaged over all the ice layers and are completely insensitive to its gradient across the various layers.
On the other hand, water ices have been estimated to have about 100-200 layers built from the beginning of the molecular cloud formation to the densest precollapse phase \citep{taquetML2012}.
These layers, hence, do not have the same [HDO]/[H$_2$O] value: the earliest layers, those closest to the refractory surfaces, are much less deuterated than those in the outer layers, formed when the temperatures were lower and the CO was also frozen  \citep[e.g.,][]{Taquet2013-HDOtheo, Caselli2012-AAR, Ceccarelli2014-PP6}.
It is therefore probable that the first layers of the water ice, where likely the bulk of the high BE sites also reside, are less deuterated, in agreement with the VSMOW value.
Another additional explanation is that the Solar System is believed to have been born in a crowded environment \citep{Adams2010-SunBirthEnvironment}, close to massive stars that heated the region.
In these regions, the molecular deuteration is lower than in the Class 0/I protostars mentioned above \citep[see, e.g., the discussion in][]{jensenHDO2021}. 
Thus, the Earth’s water deuteration ratio could align with the initial ice layers formed in moderately cold pre-stellar cores.

In summary, at the light of the above discussion, it is possible that the terrestrial water was inherited from the icy grains in the orbit of Earth. i.e. locally.
In addition, this would agree with the latest suggestions that ECs are the Earth's building blocks, formed near Earth’s orbit and possessing a D/H ratio comparable to that of the Earth's water \citep{pianiEnstatite2020}.

%%%%%%%%%%%%%%%%%%%%%%%%%%%%%%%%%%%%%%%%%%%%%%%%%%%%%%%%%%%%%%%%%%%%%%
%%%%%%%%%%%%%%%%%%%%%%%%%%%%%%%%%%%%%%%%%%%%%%%%%%%%%%%%%%%%%%%%%%%%%%
%%%%%%%%%%%%%%%%%%%%%%%%%%%%%%%%%%%%%%%%%%%%%%%%%%%%%%%%%%%%%%%%%%%%%%
\section{Conclusions} \label{sec:conclusions}

In this work, we model the distribution of water ice in the PSN using a kinetic approach based on the distribution of the water BE distribution computed by \cite{tinacciBE2023}.
We show that water sublimates gradually, on a \textit{diffuse snowline}, moving beyond the classical consideration of a sharp snowline at 180 K. 
Our model with a temperature between 145 and 200 K at 1 au successfully reproduces the highest and lowest estimates of the Earth’s water content and matches the water content trends observed across the chondrite groups at their expected formation locations. 
The derived 200 K at 1 au is consistent with the temperature measured in Class 0/I protostellar phases at these scales, when scaled for the 0.5 L$_\odot$. 
This suggests that when the luminosity of the young Class 0 Sun arrived to about 0.5 L$_\odot$ the bulk of the ice eventually inherited by the Earth was trapped inside large coagulated dust grains.

Finally, the deuteration ratio of terrestrial water is also compatible with observed ratios in protostars and the idea that water adsorbed in the highest BE sites constitutes the bulk of the terrestrial inherited water.

In summary, these results suggest that a significant share of Earth’s water could have originated locally, without requiring delivery from beyond the classical snowline.

%\textcolor{Orchid}{
%We want to emphasize here the early formation of water in cold molecular cloud. As grain growth could already begin in the very first stages of planetary formation, the icy mantle could be efficiently trapped into small bodies well before any later heating events. In our simple model, dust grains kept below \SI{200}{\kelvin} can retain enough ice to match the lower bound of Earth’s present water inventory. This threshold also relaxes the \SI{180}{\kelvin} sharp condensation front of water. Regardless of the Solar System's subsequent evolution, these results suggest that a significant share of Earth’s water could have originated locally, without requiring delivery from beyond the classical snowline.
%}

%%%%%%%%%%%%%%%%%%%%%%%%%%%%%%%%%%%%%%%%%%%%%%%%%%%%%%%%%%%%%%%%%%%%%%
%%%%%%%%%%%%%%%%%%%%%%%%%%%%%%%%%%%%%%%%%%%%%%%%%%%%%%%%%%%%%%%%%%%%%%
%%%%%%%%%%%%%%%%%%%%%%%%%%%%%%%%%%%%%%%%%%%%%%%%%%%%%%%%%%%%%%%%%%%%%%
\bibliography{Lise}{}
\bibliographystyle{aasjournal}



\end{document}